\documentclass[seceq]{ptptex}

\def\be{\begin{equation}}
\def\ee{\end{equation}}
\def\bea{\begin{eqnarray}}
\def\eea{\end{eqnarray}}
\def\Tr{{\rm tr \;}}

\title{
Two-hadron correlations in the Color Glass Condensate formalism%
}

\author{
Jamal \textsc{Jalilian-Marian}%
}

\inst{
Baruch College, 17 Lexington Ave., New York NY 10010, USA and
The Graduate School and University Center, City University of New York, 
365 Fifth Avenue, New York, NY 10016, USA
}

\abst{
Two-hadron correlations are a sensitive probe of the dynamics of gluon
saturation and the Color Glass Condensate formalism where the degrees
of freedom are Wilson lines of the gluon field. It is shown that 
unlike structure functions in DIS and single hadron production in 
proton-nucleus collisions, higher point functions of Wilson lines appear 
in two-hadron production cross section. We investigate the energy ($x$) evolution 
of these higher point functions using the JIMWLK evolution equations and 
show that dipole approximation, employed commonly in the literature to fit the
di-hadron data measured by the STAR collaboration in the forward rapidity region, breaks down. 
This necessitates an investigation of the full hierarchy of the JIMWLK 
evolution equations for these higher point functions and their solutions. 
This can then be used to make a quantitative analysis of the di-hadron correlations 
in the forward rapidity region in deuteron-gold collisions at RHIC and in the 
long range rapidity correlations observed in proton-proton collisions at the LHC.  
}

\begin{document}

\maketitle

\section{Introduction}
It is an experimental fact, established by HERA experiments on electron-proton 
collisions, that the number of gluons in the wave function of a proton grows
fast with Bjorken $x_{Bj} \sim {Q^2 \over S}$ where $Q^2$ is the virtuality of the 
exchanged photon and $S$ is the photon-hadron center of mass energy squared.
Bjorken $x_{Bj}$ can also be understood as the ratio of the energy of the proton carried by
a gluon (or any parton in general) denoted by $x$. This came as a surprise and had led 
to an explosion of ideas on the fate of gluons in the limit when 
$x \rightarrow 0$, also known as the high energy limit of QCD.

The fast growth of the gluon distribution function can be understood in
perturbative QCD to be due to sequential radiation of large number of soft (in 
longitudinal momentum) gluons which populate the wave function. In other words,
the gluon splitting function is singular in the limit $x\rightarrow 0$ 
so that due to the large longitudinal phase space available at small $x$, 
probability for gluon radiation is large even though the coupling
constant may still be small. Technically speaking, one needs to re-sum
quantum corrections which are enhanced by $ln 1/x$ which in turn leads
to a power growth of the gluon distribution function with $(1/x)^{\lambda}$
with $\lambda \sim 0.3$.

The power-like rise of the gluon distribution function would eventually
lead to a power growth of hadronic cross sections which would violate
the Froissart bound and unitarity. This fast growth however is believed
to be tamed by non-linear effects, expected to be important when the density 
of gluons in a proton is so large that the probability for gluon recombination 
becomes of the same order as the bremsstrahlung radiation, responsible for the 
fast growth of the gluon distribution function.

The gluon recombination effects were first considered by Gribov-Levin-Ryskin
in a pioneering paper~\cite{glr} where they argued that recombination diagrams
should be as important as the bremsstrahlung ones when the gluon distribution
function satisfies the following condition
\be
{\alpha_s \, xG (x, Q^2, b_t) \over S_\perp \, Q^2} \sim 1
\label{eq:sat}
\ee
where $S_\perp$ is the unit transverse area. Mueller and Qiu~\cite{mq} then
calculated the relevant diagrams in the double logarithm region and confirmed
the GLR expectation. Solving equation~(\ref{eq:sat}) 
self-consistently leads to a scale $Q_s^2 (x)$ where this relation
is satisfied. This scale is nowadays called the saturation scale 
and depends on $x$ as well as the impact parameter $b_t$. In case
of a nucleus, it will also depend on $A$, the nucleon number, as $A^{1/3}$
so that one expects that the saturation scale is large at high energy ($x\rightarrow 0$) 
and for large nuclei ($A^{1/3} \rightarrow \infty$). This means that this high gluon 
density system is weakly coupled so that one can still use weak coupling techniques 
even though the system is non-perturbative. This is a highly non-trivial observation 
first made by McLerran and Venugopalan~\cite{mv}.    

In this limit the standard expressions for particle production using collinear factorization 
in perturbative QCD breakdown due to two reasons, firstly due to the large energy effects
which appear as large logs of $1/x$ and are not re-summed in the standard expressions. Secondly,
due to the large gluon density the standard twist expansion is not valid anymore in the sense
that all twist terms are as large as the leading twist term. This necessitates introduction 
of another formalism, capable of including both effects. The formalism which generalizes 
perturbative QCD to include both of these effects has come to be known the Color Glass 
Condensate formalism and the high gluon density effects are referred to as gluon saturation.

\section{The Color Glass Condensate}

In order to go beyond the leading-twist perturbative QCD and to include both large
logs of energy and gluon density, McLerran and Venugopalan introduced an effective action
which treats the large $x$ degrees of freedom as color charges $\rho$ to which the gluon
fields $A^{\mu}$ couple. Due to the small $x$ approximation, the coupling between the 
color charges $\rho$ and the gluon field is assumed to be eikonal and given by a Wilson
line of the gauge field $A_\mu$. The distribution 
of color charges $\rho$ is non-perturbative and assumed to be given by a weight functional
$W[\rho]$. To calculate an observable, one solves the classical equations of motion for a
fixed color charge $\rho$ and then averages over all color charges $\rho$ with the weight 
functional $W[\rho]$. The weight functions $W[\rho]$ satisfies the so-called JIMWLK evolution
equation~\cite{jimwlk} which describes the evolution of $W$ with energy or equivalently with 
rapidity $y$ ($y = log\, 1/x$).
The JIMWLK evolution equation for rapidity evolution of any operator $O$ can be written as
\be
\frac{d}{dy} \langle O \rangle =
  \frac{1}{2} \left< \int d^2x \, d^2y \, \frac{\delta}{\delta\alpha_x^b}
     \, \eta^{bd}_{xy} \, \frac{\delta}{\delta\alpha_y^d} \, O \right>~,
   \label{eq:ham}
\ee
where
\be  \label{eq:eta}
\eta^{bd}_{xy} = \frac{1}{\pi} \int \frac{d^2z}{(2\pi)^2}
     \frac{(x-z)\cdot(y-z)}{(x-z)^2 (y-z)^2} \left[
       1 + U^\dagger_x U_y - U^\dagger_x U_z - U^\dagger_z U_y
       \right]^{bd} 
\ee
and $U_x \equiv U (x_t)$ is the Wilson line in the adjoint representation. This formalism
has been applied to any high energy process which involves at least one proton or nucleus
in the initial state. Examples are Deeply Inelastic Scattering (DIS) of electrons
on protons and nuclei, and particle production in hadronic/nuclear collisions. The simplest 
process to consider is DIS structure functions which are related to the total (virtual) photon-hadron
cross section. In the rest frame of the target, this process can be factored into two parts; first
the virtual photon splits into a quark anti-quark dipole which then scatters on the target. The
probability for the virtual photon to split into a quark anti-quark dipole is given by the square
of the photon wave function and is calculable in QED. The subsequent scattering of the quark anti-quark
dipole on the target is described by the CGC formalism which describes its energy (rapidity or $x$) 
dependence. The JIMWLK evolution equation for the rapidity evolution of the quark anti-quark dipole 
can be written as
\be
{d\over dy} \langle \Tr V_r^\dagger \, V_s \rangle = -
{N_c\, \alpha_s \over 2\pi^2} \int\,
d^2 z\, {(r - s)^2 \over (r - z)^2 (s - z)^2} \, \left<
\Tr V_r^\dagger \, V_s - {1\over N_c} \Tr V_r^\dagger \, V_z~
\Tr\, V_s \, V^\dagger_z\right>
\label{eq:2pt}
\ee 
where $r, s$ are transverse coordinates of the quark and anti-quark. 
The left hand side of the equation describes the rapidity dependence of the probability 
for scattering of a fundamental (quark anti-quark) dipole on the target proton or nucleus.
Due to the non-linear term on the right hand side, this probability is unitary in the sense 
that it can never be larger than one, unlike probabilities calculated in leading-twist pQCD. 
There are, however, no known analytic solutions to this equation and one has to resort to
approximate methods in order to gain further understanding of its properties. One very common
approximation is the leading $N_c$ and dipole approximation such that the color average (denoted
by the brackets in the above equation) of the product of any number of Wilson lines is replaced 
by products of color averages of two Wilson lines. With this approximation, the JIMWLK
equation for the fundamental dipole ($2$-pt function) reduces to the Balitsky-Kovchegov (BK) 
equation~\cite{bk}. Normalizing the $2$-pt function as
\be 
S (r - s) \equiv {1\over N_c}  < \Tr V^\dagger_r\,  V_s >
\label{eq:S2_def}
\ee
the BK equation for the evolution of a fundamental dipole is written as 
\be
{d\over dy} S (r - s)  = 
- {N_c\, \alpha_s \over 2\pi^2} \;
\int d^2 z \,
 {(r - s)^2 \over (r - z)^2 (s - z)^2}\,\bigg[S (r -s) -
  S (r -z) \, S (z - s) \bigg]
\label{eq:bk}
\ee
The dipole profile function 
$S (r - s)$ is the building block for many observables such
as DIS structure functions at small $x$ and forward particle production in proton-nucleus 
collisions at high energy. Quite recently the next-to-leading order corrections to the BK 
equation have been derived and the running coupling solution is obtained numerically~\cite{nlo}. 
The single inclusive hadron production cross section is given by~\cite{adjjm1,dhj} 
\bea
{d\sigma^{p A \rightarrow h X} \over dY \, d^2 P_t \, d^2 b} &=& 
{1\over (2\pi)^2}
\int_{x_F}^{1} dx \, {x\over x_F} \Bigg\{
f_{q/p} (x,Q^2)~ S [{x\over x_F} P_t , b]~ D_{h/q} ({x_F\over
  x}, Q^2) +  \nonumber \\
&&
f_{g/p} (x,Q^2)~ S_A [{x\over x_F} P_t , b] ~ D_{g/h} ({x_F\over x}, Q^2) 
\Bigg\}
\label{eq:single}
\eea
where $S$ and $S_A$ are the fundamental and adjoint dipoles. Using phenomenological
models of the dipole profiles above has led to a quantitative description of the RHIC
data on forward rapidity hadron production in dA collisions (the result for $y=4$ was 
a prediction~\cite{dhj} which was later confirmed experimentally). Most recently the solution to the 
running coupling BK equation has been used to describe the data successfully~\cite{jayk} which gives
one more confidence about the applicability of the CGC formalism in the forward rapidity 
region of RHIC. Nevertheless, models based on the standard collinear factorization
can also describe the forward rapidity data by including shadowing and cold matter energy
loss\cite{ivan}. Measurement of other processes, for instance photon and dilepton production in the forward
rapidity region~\cite{em} will help clarify the dynamics of the forward rapidity particle production 
(see~\cite{review} for a review of CGC and its applications to particle production in proton-nucleus collisions).

\subsection{Two-particle correlations}
Two-particle correlations are expected to contain more information about the dynamics of
the process than single inclusive production. In addition to transverse momentum
dependence of the cross section, one can investigate the angular dependence of the cross
section which can shed more light on the dynamics of the process. For instance, in the
standard collinear factorization approach to two hadron production, the produced partons
are back to back (in Leading Order) so that one expects a sharp peak on the away side ($\pi$).
On the other hand, in the CGC formalism, one expects to have a disappearance of the peak, 
due to shadowing generated by small $x$ re-summation and $p_t$ broadening due to multiple 
scattering. The simplest examples of two-parton productions in the CGC framework include 
quark anti-quark ~\cite{fgjjm} and $2$-gluon~\cite{ykjjm1} production in DIS. 

Two-hadron angular correlations in deuteron-gold collisions in the forward rapidity region have 
been recently measured by the STAR collaboration at RHIC. For central collisions a disappearance
of the away side hadron is observed as expected in the CGC formalism. The underlying partonic
process is a projectile quark scattering from the target nucleus and radiating a gluon either
before or after the scattering. The expressions for this process are given in~\cite{ykjjm1}
in momentum space. Later they were also derived in the coordinate space which leads to a more
compact form of the equations~\cite{cm}. We refer the reader to~\cite{ykjjm1, cm} for the 
details of derivation and the explicit form of the production cross section. Here we just note
that production of a quark and gluon in dA collisions involves higher point (more than $2$) 
functions of Wilson lines which were not present in DIS structure functions or single inclusive
hadron production in dA collisions. For instance, the following products of Wilson lines, denoted
$O_4$ and $O_6$ appear in the production cross section
\be
O_4 (r, \bar{r}:s) \equiv \Tr V^\dagger_r\, t^a \, V_{\bar{r}}\, t^b
\, [U_s]^{ab}  = 
{1\over 2} \bigg[ \Tr V_r^\dagger\, V_s ~\, \Tr V_{\bar{r}} \, V_s^\dagger - 
{1\over N_c} \Tr V_r^\dagger \, V_{\bar{r}} \bigg]
\label{eq:o_4}
\ee
and 
\be
O_6 (r,\bar{r}:s,\bar{s}) \equiv \Tr V_r \, V^\dagger_{\bar{r}}\,
t^a\, t^b \, [U_s\, U^\dagger_{\bar{s}}]^{ba} =
{1\over 2} \bigg[
\Tr V_r \, V^\dagger_{\bar{r}}\, V_{\bar{s}} \, V^\dagger_s
~ \, \Tr V_s \, V^\dagger_{\bar{s}} 
- {1\over N_c} \Tr V_r \, V^\dagger_{\bar{r}}\bigg]
\label{eq:o_6}
\ee
where the following identity is used
\be
U^{ab}\, t^b = V^\dagger \, t^a\, V.
\label{eq:a2f}
\ee
In principle one will need to know the equation describing the evolution of 
these operators with rapidity, the same way that the JIMWLK-BK equations describes
the rapidity evolution of the weight function and the $2$-pt function. The leading
$N_c$ part of the equation describing the rapidity evolution of $4$ Wilson lines is
derived in~\cite{ykjjm1} while the complete evolution equation for the operators 
$O_4$ and $O_6$ (which appear in the di-jet production) is derived in~\cite{higher1}. 
Here we just write the results and refer the reader to~\cite{higher1} for the details. 
The evolution equation for the $4$-pt function is 
\bea
&&{d\over dy} \langle O_4 (r,\bar{r}:s) \rangle  =  
- {N_c\, \alpha_s \over (2\pi)^2} 
\int d^2 z \, \Biggl< 
2\, \bigg[{(r - s)^2 \over (r - z)^2 (s - z)^2} + {(\bar{r} - s)^2 \over
    (\bar{r} - z)^2 (s - z)^2 }\bigg] \,  O_4 (r,\bar{r}:s) \, 
\nonumber \\
&-& {1\over N_c} \Bigg[
{(r - s)^2 \over (r - z)^2 (s - z)^2}\, \Tr V_r^\dagger \, V_z ~\,
\Tr  V^\dagger_s\, V_{\bar{r}} ~\, \Tr V_z^\dagger \, V_s 
+ \nonumber \\
&&{(\bar{r} - s)^2 \over (\bar{r} - z)^2 (s - z)^2 }
\Tr V_r^\dagger \, V_s ~\, \Tr V^\dagger_z\, V_{\bar{r}} ~\, 
\Tr\, V_s^\dagger \, V_z \, - 
\nonumber \\
&&
{1\over 2} \, 
\bigg[ {(r - s)^2 \over (r - z)^2 (s - z)^2} + 
{(\bar{r} - s)^2 \over (\bar{r} - z)^2 (s - z)^2 }
- {(r - \bar{r})^2 \over (r - z)^2 (\bar{r} - z)^2 } \bigg] \nonumber \\
&& 
\bigg[\Tr V_r^\dagger \, V_z \, V^\dagger_s\, V_{\bar{r}} \, V_z^\dagger \, V_s +  
\Tr V_r^\dagger \, V_s \, V^\dagger_z\, V_{\bar{r}} \, V_s^\dagger \,
V_z\bigg] 
\Bigg]
\nonumber \\
&+&  {1\over N_c^2} \, {(r - \bar{r})^2 \over (r - z)^2 (\bar{r} -
  z)^2} \, \Tr\, V_r^\dagger \, V_z  ~\, \Tr\, V_z^\dagger \, V_{\bar{r}}
\;\Biggr >
\label{eq:ev_o_4}
\eea
while the evolution equation for the $6$-pt function can be written as
\bea
&&{d\over dy} \langle O_6 (r,\bar{r}:s,\bar{s}) \rangle
 =  - {N_c\, \alpha_s \over 2(2\pi)^2} \int d^2 z \Bigg<  
2\, \big[
{(r - s)^2 \over (r - z)^2 (s - z)^2} +  
{(r - \bar{r})^2 \over (r - z)^2 (\bar{r} - z)^2} +  \nonumber \\
&&
{(\bar{r} - \bar{s})^2 \over (\bar{r} - z)^2 (\bar{s} - z)^2}
+   3 {(s - \bar{s})^2 \over (s - z)^2 (\bar{s} - z)^2}\big] \, 
O_6 (r,\bar{r}:s,\bar{s}) - {1\over N_c} \Bigg[\nonumber\\
&&
\big[ {(r - \bar{r})^2 \over (r - z)^2 (\bar{r} - z)^2} + {(r - s)^2
    \over (r - z)^2 (s - z)^2} -
{(s - \bar{r})^2 \over (s - z)^2 (\bar{r} - z)^2}\big] \,
\Tr V_z \, V^\dagger_{\bar{r}} \, V_{\bar{s}} \, V^\dagger_s ~\, 
\Tr V_r \, V^\dagger_z ~\,
\Tr V_s \, V^\dagger_{\bar{s}} \nonumber \\
&&
+ \big[ {(r - \bar{r})^2 \over (r - z)^2 (\bar{r} - z)^2} + {(\bar{r}
    - \bar{s})^2 \over (\bar{r} - z)^2 (\bar{s} - z)^2} - {(r -
    \bar{s})^2 \over (r - z)^2 (\bar{s} - z)^2}\big]\,
\Tr V_r \, V^\dagger_z \, V_{\bar{s}} \, V^\dagger_s ~\, 
\Tr V_z \, V^\dagger_{\bar{r}} ~\, 
\Tr V_s \, V^\dagger_{\bar{s}} \nonumber \\
&&     
+ \big[ {(r - s)^2 \over (r - z)^2 (s - z)^2} + {(s - \bar{s})^2
    \over (s - z)^2 (\bar{s} - z)^2} - {(r - \bar{s})^2 \over (r -
    z)^2 (\bar{s} - z)^2}\big]\,
\Tr V_r \, V^\dagger_{\bar{r}} \, V_{\bar{s}} \, V^\dagger_z ~\, 
\Tr V_z \, V^\dagger_s ~\, 
\Tr V_s \, V^\dagger_{\bar{s}} \nonumber \\
&&     
+ \big[ {(\bar{r} - \bar{s})^2 \over (\bar{r} - z)^2 (\bar{s} - z)^2}
  + {(s - \bar{s})^2 \over (s - z)^2 (\bar{s} - z)^2} - {(\bar{r} -
    s)^2 \over (\bar{r} - z)^2 (s - z)^2}\big]\,
\Tr V_r \, V^\dagger_{\bar{r}} \, V_z \, V^\dagger_s ~\, 
\Tr V_{\bar{s}} \, V^\dagger_z ~\, 
\Tr V_s \, V^\dagger_{\bar{s}} \nonumber \\
&&     
+ 2 {(s - \bar{s})^2 \over (s - z)^2 (\bar{s} - z)^2} \,
\Tr V_r \, V^\dagger_{\bar{r}} \, V_{\bar{s}} \, V^\dagger_s ~\, 
\Tr V_s \, V^\dagger_z \, 
\Tr V_z \, V^\dagger_{\bar{s}} \nonumber \\
&&    
+ \big[{(\bar{r} - s)^2 \over (\bar{r} - z)^2 (s - z)^2} - 
{(\bar{r} - \bar{s})^2 \over (\bar{r} - z)^2 (\bar{s} - z)^2}
- {(r - s)^2 \over (r - z)^2 (s - z)^2} +  
{(r - \bar{s})^2 \over (r - z)^2 (\bar{s} - z)^2} \big]\nonumber \\
&&
\Tr V_r \, V^\dagger_s ~\, 
\Tr\, V^\dagger_{\bar{r}}\, V_{\bar{s}} ~\, 
\Tr V_s \, V^\dagger_{\bar{s}} \nonumber \\
&&     
-  \big[- {(\bar{r} - s)^2 \over (\bar{r} - z)^2 (s - z)^2} - {(r -
    \bar{s})^2 \over (r - z)^2 (\bar{s} - z)^2}
+ {(r - \bar{r})^2 \over (r - z)^2 (\bar{r} - z)^2} +  {(s -
  \bar{s})^2 \over (s - z)^2 (\bar{s} - z)^2} \big]\nonumber \\
&&
\Tr V_r \, V^\dagger_{\bar{r}} \, 
\Tr V_{\bar{s}}\, V^\dagger_s ~\, 
\Tr V_s \, V^\dagger_{\bar{s}} \nonumber \\
&&   
- \big[{(r - s)^2 \over (r - z)^2 (s - z)^2} - 
{(r - \bar{s})^2 \over (r - z)^2 (\bar{s} - z)^2} + 
{(s - \bar{s})^2 \over (s - z)^2 (\bar{s} - z)^2}\big] 
\Tr V_r \, V^\dagger_{\bar{r}}\, V_{\bar{s}} \, V^\dagger_s\,  V_z \,
V^\dagger_{\bar{s}}\, V_s \, V^\dagger_z \nonumber \\
&&
+  \big[ {(\bar{r} - s)^2 \over (\bar{r} - z)^2 (s - z)^2} -
  {(\bar{r} - \bar{s})^2 \over (\bar{r} - z)^2 (\bar{s} - z)^2} -
{(s - \bar{s})^2 \over (s - z)^2 (\bar{s} - z)^2}\big] 
\Tr V_r \, V^\dagger_{\bar{r}}\, V_z \, V^\dagger_{\bar{s}} \, V_s\,
V^\dagger_z \, V_{\bar{s}}\, V^\dagger_s
\nonumber\\
&& +  \big[{(\bar{r} - s)^2 \over (\bar{r} - z)^2 (s - z)^2} -
  {(\bar{r} - \bar{s})^2 \over (\bar{r} - z)^2 (\bar{s} - z)^2} - 
{(r - s)^2 \over (r - z)^2 (s - z)^2} +  
{(r - \bar{s})^2 \over (r - z)^2 (\bar{s} - z)^2} \big]\,\nonumber \\
&&
\Tr V_r \, V^\dagger_z\, V_s \, V^\dagger_{\bar{s}} \, V_z\,
V^\dagger_{\bar{r}} \, V_{\bar{s}}\, V^\dagger_s 
- 2 {(s - \bar{s})^2 \over (s - z)^2 (\bar{s} - z)^2} \,
\Tr V_r \,V^\dagger_{\bar{r}}\, V_{\bar{s}} \, V^\dagger_z \, V_s\,
V^\dagger_{\bar{s}} \, V_z\, V^\dagger_s
\nonumber \\
&& 
- \big[{(\bar{r} - s)^2 \over (\bar{r} - z)^2 (s - z)^2} - {(\bar{r} -
    \bar{s})^2 \over (\bar{r} - z)^2 (\bar{s} - z)^2} - {(r - s)^2
    \over (r - z)^2 (s - z)^2} +  {(r - \bar{s})^2 \over (r - z)^2
    (\bar{s} - z)^2} \big]\nonumber \\
&&
\Tr V_r \, V^\dagger_{\bar{s}} \, V_s\,  V^\dagger_{\bar{r}} \,
V_{\bar{s}}\, V^\dagger_s
\nonumber\\
&&
- \big[{(\bar{r} - s)^2 \over (\bar{r} - z)^2 (s - z)^2} + 
{(r - \bar{s})^2 \over (r - z)^2 (\bar{s} - z)^2} 
- {(s - \bar{s})^2 \over (s - z)^2 (\bar{s} - z)^2} \big]\,
\Tr V_r \, V^\dagger_{\bar{r}}\Bigg] \nonumber \\
&+&
{1\over N_c^2}  {(r - \bar{r})^2 \over (r - z)^2 (\bar{r} - z)^2} \,
\Tr V^\dagger_r\, V_z ~\, \Tr V^\dagger_z\,  V_{\bar{r}} \Bigg> 
- {1\over 4\, N_c} \, {d\over d y}\, <\Tr V^\dagger_r \, V_{\bar{r}}>.
\label{eq:ev_o_6}
\eea
We note that these equations are free of power divergences in the limit 
the internal transverse coordinate approaches any of the external
coordinates. Unlike the evolution equation for the dipole profile ($2$-pt function)
which is rather simple, these equations are quite involved and it is not easy to
investigate their properties analytically (see~\cite{fabio} for particular kinematics
in which these relations get a bit simplified is investigated). In phenomenological 
applications of CGC 
to two-hadron production it is assumed~\cite{jacm} that in the leading $N_c$ 
approximation one can write the product of higher point functions as the products 
of two point functions only (the dipole approximation):
\be
\langle O_6 (r,\bar{r}:s,\bar{s}) 
\rangle \; \simeq \; 
\langle O_2 (r - s) \rangle \; \langle O_2 (\bar{r} - \bar{s}) \rangle \; \langle O_2
(s - \bar{s}) \rangle  \, +
\langle O_2 (r - \bar{r}) \rangle \; \langle O_2 (\bar{s} - s) \rangle \; \langle O_2
(s - \bar{s}) \rangle. 
\label{eq:o_dipole}
\ee 
This assumption has been made by Albacete and Marquet~\cite{jacm} in order to fit the data on forward 
rapidity di-jet correlations in dA collisions at RHIC. Unfortunately, {\it this (dipole) assumption is 
wrong} and misses many leading $N_c$ contributions to the evolution equation as 
proven in~\cite{higher1}. Therefore, a quantitative understanding of the forward rapidity two-hadron
correlations from CGC is still lacking. 

There are several issues that need to be understood. First, we note that the number ($n$) of terms 
in the evolution equations for the higher point functions which are formally $N_c$ suppressed 
becomes very large so that when $n >> N_c$, large $N_c$ approximation may cease to be a good 
approximation. To see this, it is convenient to define normalized operators $S_6$ and $S_4$ such that 
\bea
S_4 (r, \bar{r}: s)&\equiv& {1\over C_A\, C_F} \, \langle O_4 \rangle
\nonumber \\
S_6 (r, \bar{r}: s, \bar{s}) &\equiv& {1\over C_A\, C_F} \, \langle
O_6 \rangle~.
\label{eq:norm_s}
\eea
Here we focus on $S_6$ since it is the more interesting one. Rewriting the evolution
for $O_6$ in terms of $S_6$, we note that the terms in the first few lines in 
eq. (\ref{eq:ev_o_6}) which are products of three traces will end up being leading 
order in $N_c$ while the next few lines which involve only one trace will be suppressed
by $N_c^2$. We note the number of $N_c^2$ suppressed terms is $8$ for $O_4$ and $\sim 20$
for $O_6$. Making a Gaussian approximation (see \cite{bgv}) to these higher point functions
would result in further proliferation of these $N_c$ suppressed terms.

The second point that needs to be understood better is the energy dependence of expectation
value of trace of a large number of Wilson lines that appear on the right hand side of 
eq. (\ref{eq:ev_o_6}). One may expect that these higher point functions will grow
faster with energy than the two-point function. There is an example where this happens,
energy dependence of a state of four reggeized gluons has been investigated 
in~\cite{am} where it is found that it has a faster rate of growth with energy
than the state of two reggeized gluons. Whether a similar thing happens in our case
is not known for sure but is likely to be true. This would mean that the terms with
larger number of Wilson lines on the right hand side of evolution equation will grow
faster with energy than the other terms which have fewer Wilson lines. This stronger
energy dependence may eventually compensate for the $N_c^2$ suppression.

Furthermore we note the presence of terms on the right hand side of eq.~(\ref{eq:ev_o_6}) 
which involve lower point functions, namely the $2$-pt and $4$-pt functions, reminiscent of 
pomeron loop contribution to the BK equation for the dipole profile~\cite{pomloop}. We note
that these terms are in addition to the original $2$-pt function which was present in the definition
of $O_6$. The origin of these terms seems to be due to kinematics but needs to
be better understood. Therefore, di-jet production in the forward rapidity region of 
deuteron-gold collisions offer a rich context and a unique opportunity to investigate 
CGC correlations. For this, one needs to solve the full JIMWLK equation for the weight 
function $W[\rho]$ which can then be used to compute any $n$-pt function in the CGC formalism.

We note that
photon-hadron correlations~\cite{phohad} in the forward rapidity region in deuteron-gold collisions is another
process which is more sensitive to saturation dynamics than single inclusive production. It also 
has the advantage, compared to two-hadron production, that it is sensitive only to the $2$-pt function
so that one could use the latest results for running coupling BK equation in order to make quantitative
predictions.

Another example of the importance of higher point functions in the CGC framework is the two-hadron
production process in proton-proton or nucleus-nucleus collisions. In this case one can not solve
the problem analytically in the full kinematics of CGC and has to resort to numerical methods.
Nevertheless, it is common to use the so-called $k_t$ factorization in the dilute region, i.e., 
where both the target and projectile are dilute but the energy of the collision is high enough
so that a re-summation in $x$ is required (the BFKL region). In this case the two-gluon production
cross section is given by~\cite{dgmv} 
\bea
&&\left\langle \frac{dN_2}{d^2pdy_p\,d^2qdy_q} \right\rangle  =
\frac{g^{12}}{64 (2\pi)^6}\, 
\left(
f_{g aa^\prime} f_{g^\prime b b^\prime}
f_{gcc^\prime} f_{g^\prime dd^\prime} 
\right)
\int \prod_{i=1}^4 \frac{d^2 k_{i}}{(2\pi)^2 k_{i}^2} \nonumber \\
&&
\frac{L_\mu(p,k_{1})L^\mu(p,k_{2})}{(p-k_{1})^2(p-k_{2})^2}
\frac{L_\nu(q,k_{3})L^\nu(q,k_{4})}{(q-k_{3})^2 (q-k_{4})^2}
\left< {\rho^*}_A^a (k_{2}) {\rho^*}_A^{b} (k_{4}) 
{{\rho}_A}^c (k_{1}) {{\rho}_A}^d (k_{3})\right> \nonumber \\
&&
\quad \times \left<
{\rho^*}_B^{a^\prime} (p-k_{2})
{\rho^*}_B^{b^\prime} (q-k_{4})
{{\rho}_B}^{c^\prime}(p-k_{1}) 
{{\rho}_B}^{d^\prime} (q-k_{3})\right> 
\label{eq:2glue}
\eea 
where $p, y_p$, $q, y_q$ are the transverse momenta and rapidities 
of the two produced gluons, $A$ and $B$ label the color charge $\rho$
of the projectile and target proton or nucleus and the Lipatov
vertex is denoted by $L^\mu$. The color charge $\rho$ is the source
for the classical gluon field which satisfies the relation (in the
covariant gauge)
\be
A^\mu(x^+,r) \equiv \delta^{\mu -} \alpha(x^+,r) = 
     - g \, \delta^{\mu -} \delta(x^+) \frac{1}{\nabla^2_\perp} \rho(x^+,r)~,
\ee
The standard un-integrated gluon distribution function $\Phi (x, p_t^2)$ 
is defined in terms of the color average of two $\rho$'s as
\be 
\label{eq:intdist}
\left< {\rho^*}^a(k) \rho^b(k')\right> (x) = \frac{1}{\alpha_s} 
   \frac{\delta^{ab}}{N_c^2-1} (2\pi)^3 \delta(k-k') \, \Phi(x,k^2)
\ee
Due to the complexity of the expression, it is common~\cite{dgmv} to make the dipole
approximation and write (symbolically) 
\be 
\label{eq:4rho_dipole}
\left< \rho^a \rho^b \rho^c \rho^d \right> =
 \delta^{ab} \delta^{cd} (\rho^2)^2 + 
 \delta^{ac} \delta^{bd} (\rho^2)^2 + 
 \delta^{ad} \delta^{bc} (\rho^2)^2 + \cdots
\ee
where $\rho^2$ denotes $\langle\rho\rho\rangle$. Again, the dipole approximation
breaks down under rapidity evolution. The correct evolution equation for the product
of four $\rho$'s was derived in~\cite{higher2} and reads
\bea
\frac{d}{dY} && 
\langle \alpha_r^a \alpha_{\bar r}^b \alpha_s^c \alpha_{\bar s}^d \rangle = 
\frac{g^2N_c}{(2\pi)^3} \int d^2z \nonumber \\
&&
\left<
  \frac{\alpha_z^a \alpha_{\bar r}^b \alpha_s^c \alpha_{\bar s}^d}{(r-z)^2} +
  \frac{\alpha_r^a \alpha_z^b \alpha_s^c \alpha_{\bar s}^d}{(\bar{r}-z)^2} +
  \frac{\alpha_r^a \alpha_{\bar r}^b \alpha_z^c \alpha_{\bar s}^d}{(s-z)^2} +
  \frac{\alpha_r^a \alpha_{\bar r}^b \alpha_s^c \alpha_z^d}{(\bar{s}-z)^2}
  -4 \frac{\alpha_r^a \alpha_{\bar r}^b \alpha_s^c \alpha_{\bar s}^d}{z^2}
\right> \nonumber\\
&& 
+ \frac{g^2}{\pi} \int \frac{d^2z}{(2\pi)^2} \nonumber \\
&&
\left<
 f^{e\kappa a}f^{f\kappa b} \frac{(r-z)\cdot(\bar r-z)}{(r-z)^2 (\bar r-z)^2} 
 \left[\alpha_r^e\alpha_{\bar r}^f - \alpha_r^e\alpha_z^f - 
       \alpha_z^e\alpha_{\bar r}^f + \alpha_z^e\alpha_z^f \right]
       \alpha_s^c \alpha_{\bar s}^d \right. \nonumber\\
& & +
 f^{e\kappa a}f^{f\kappa c} \frac{(r-z)\cdot(s-z)}{(r-z)^2 (s-z)^2} 
 \left[\alpha_r^e\alpha_s^f - \alpha_r^e\alpha_z^f - 
       \alpha_z^e\alpha_s^f + \alpha_z^e\alpha_z^f \right]
       \alpha_{\bar r}^b \alpha_{\bar s}^d \nonumber\\
& & +
 f^{e\kappa a}f^{f\kappa d} \frac{(r-z)\cdot(\bar s-z)}{(r-z)^2 (\bar s-z)^2} 
 \left[\alpha_r^e\alpha_{\bar s}^f - \alpha_r^e\alpha_z^f - 
       \alpha_z^e\alpha_{\bar s}^f + \alpha_z^e\alpha_z^f \right]
       \alpha_{\bar r}^b \alpha_s^c \nonumber\\
& & +
 f^{e\kappa b}f^{f\kappa c} \frac{(\bar r-z)\cdot(s-z)}{(\bar r-z)^2 (s-z)^2} 
 \left[\alpha_{\bar r}^e\alpha_s^f - \alpha_{\bar r}^e\alpha_z^f - 
       \alpha_z^e\alpha_s^f + \alpha_z^e\alpha_z^f \right]
       \alpha_r^a \alpha_{\bar s}^d \nonumber\\
& & +
 f^{e\kappa b}f^{f\kappa d} \frac{(\bar r-z)\cdot(\bar s-z)}{(\bar r-z)^2 (\bar s-z)^2} 
 \left[\alpha_{\bar r}^e\alpha_{\bar s}^f - \alpha_{\bar r}^e\alpha_z^f - 
       \alpha_z^e\alpha_{\bar s}^f + \alpha_z^e\alpha_z^f \right]
       \alpha_r^a \alpha_s^c \nonumber\\
& & + \left.
 f^{e\kappa c}f^{f\kappa d} \frac{(s-z)\cdot(\bar s-z)}{(s-z)^2 (\bar s-z)^2} 
 \left[\alpha_s^e\alpha_{\bar s}^f - \alpha_s^e\alpha_z^f - 
       \alpha_z^e\alpha_{\bar s}^f + \alpha_z^e\alpha_z^f \right]
       \alpha_r^a \alpha_{\bar r}^b \right>~.  
\label{eq:ev_4rho}
\eea
This equation is valid in the dilute region of the proton or nucleus so that
all higher order terms in the field $\alpha$ have been neglected on the right hand side
of the equation. Solving this equation (which can be done using numerical methods), one
can use the solution in the expression for two-gluon production~(\ref{eq:2glue}) and investigate
the dependence of the production cross section on the transverse momenta and rapidities 
of the two gluons. This will be extremely interesting due to the recent results from heavy ion
collisions at RHIC and proton-proton collisions at LHC where long range rapidity correlations
are observed~\cite{cms}. These long range rapidity correlations must be generated very early 
after the collision and subsequent re-scattering can not change these due to causality~\cite{dgmv}. 
Furthermore, in proton-proton collisions one does not expect to create a medium with a size and 
life time much larger than a Fermi, which is how the heavy ion community defines a Quark-Gluon 
Plasma. Thus, viscous corrections to ideal hydrodynamics must necessarily be {\it very large} for
transverse momenta larger than a few hundred $MeV$. Therefore, one does not expect to see flow
effects above one $1 \, GeV$. 

In the CGC formalism long range rapidity correlations are due to the production of
boost-invariant longitudinal color fields at the very early stages of the collision. These classical 
longitudinal fields are boost invariant which leads to production of gluons independent of
rapidity. Furthermore, the transverse size (correlation length) of these fields is inversely
proportional to the saturation momentum $Q_s$ and decreases with increasing collision energy.
This means that most produced particles will have transverse momenta of the order of $Q_s$
and will be produced approximately independently of rapidity. These features of the data from 
RHIC and LHC are in qualitative agreement with the expectations from the CGC formalism~\cite{ridge_cgc}. 
Nevertheless, a truely quantitative  comparison with the data requires knowledge of the higher 
point functions in CGC formalism.
 
\section*{Acknowledgements}
We would like to thank F. Dominguez and A.H. Mueller for interesting discussions. 
The results presented here are based on work in collaboration with A. Dumitru and have been
presented by the author in the program "High Energy Strong Interactions 2010" at the Yukawa 
Institute for Theoretical Physics, July-August 2010, Kyoto, Japan and in the "5th International 
Workshop on high $p_t$ Physics at LHC 2010", September - October, 2010, Mexico City, Mexico. 
This work is supported by the DOE Office of Nuclear Physics
through Grant No.\ DE-FG02-09ER41620 and by The City University of
New York through the PSC-CUNY Research Award Program, grant number 
62625-40.

%

\end{document}